\documentclass[journal,twocolumn]{IEEEtran}

\usepackage{cite}
\usepackage{amsmath,amsfonts,amsxtra,amssymb,latexsym,amscd,amsthm,mathrsfs,bm}

\usepackage{graphicx}

\usepackage{hyperref}
\usepackage{xcolor}
\usepackage{algorithm}
\usepackage{algorithmic}
\usepackage[small]{caption}
\usepackage{subcaption}
\usepackage[colorinlistoftodos,textsize=tiny]{todonotes}
\usepackage{lipsum}
\usepackage{cleveref}

\crefname{figure}{}{}
\Crefname{figure}{}{}

\usepackage{soul}
\makeatletter
\if@todonotes@disabled

\else

\fi
\makeatother

\usepackage{stfloats}

\usepackage[letterpaper, top=0.75in, bottom=1in, left=0.625in, right=0.625in]{geometry}

\usepackage{epstopdf}
\title{Index Modulation for Modulation on Conjugate-Reciprocal Zeros (IM-MOCZ)}
\author{Aidan Corbett and Ebrahim Bedeer
\thanks{A. Corbett and E. Bedeer are with the Department of Electrical and Computer Engineering,
		University of Saskatchewan, Saskatoon, SK, Canada S7N 5A9. Emails: \{aidan.corbett, e.bedeer\}@usask.ca}
}
\IEEEaftertitletext{\vspace{-1.2\baselineskip}}

\begin{document}
\maketitle
\thispagestyle{empty}
\pagestyle{empty}

\begin{abstract}
This paper investigates the application of Index Modulation (IM) to Modulation on Conjugate-Reciprocal Zeros (MOCZ) to enhance spectral efficiency (SE) in short packet communications. The proposed IM-MOCZ scheme splits an $N$-bit message into two streams: $N-K$ bits select one of $2^{N-K}$ uniquely designed codebooks, while the remaining $K$ bits are transmitted with conventional binary-MOCZ (BMOCZ) using the selected codebook. At the receiver, Root Finding Minimum Distance (RFMD) or Direct Zero-Testing (DiZeT) detectors evaluate all candidate codebooks and compute penalty metrics, with a majority vote rule selecting the most confident codebook and recovering the transmitted message. The proposed IM-MOCZ can provide higher SE gains than conventional BMOCZ at the cost of increasing the computational complexity, with simulations demonstrating improved bit error rate (BER) and block error rate (BLER) performance for larger $K$ relative to $N$, when compared to conventional BMOCZ.
\end{abstract}


\vspace{-0.5em}
\section{Introduction}\label{sec:Introduction}

\begin{figure*}[t]	
	\centering
	{\includegraphics[width=\textwidth]{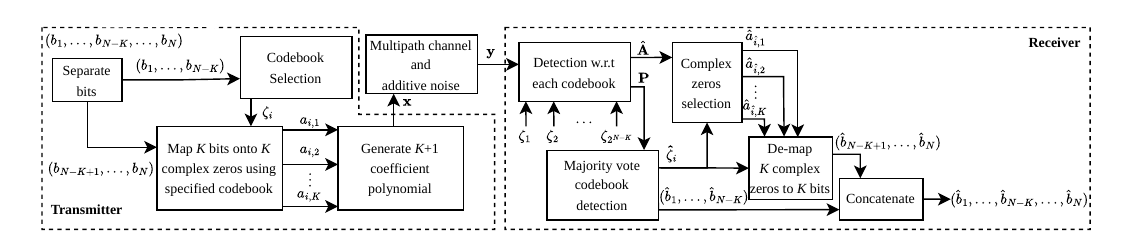}}
	\caption{A block diagram of the proposed baseband equivalent IM-MOCZ system.}
    \vspace{-1em}
	\label{fig:IM_MOCZ_BlockDiagram}
\end{figure*}

Emerging wireless applications such as ultra-reliable low-latency communication (URLLC) and the Internet of Things (IoT) require the reliable transmission of short packets under strict latency and reliability constraints. In short-packet communications (SPCs), traditional coherent methods become inefficient, as the overhead required for channel estimation can rival or exceed the payload itself, significantly degrading spectral efficiency (SE) and system performance. Modulation on Conjugate-Reciprocal Zeros (MOCZ) addresses this challenge as a non-coherent modulation scheme specifically designed for SPC scenarios \cite{MOCZ}. By eliminating the need for pilot-based channel estimation, MOCZ reduces overhead while maintaining reliable detection, making it particularly attractive for sporadic, low-latency transmissions.

Since its introduction, MOCZ has been extended in several directions. Walk et al. 
established the foundational framework and its suitability for non-coherent SPC 
\cite{MOCZ}, and later explored multi-user extensions \cite{MU-MOCZ}. Siddiqui et al. improved SE 
by integrating faster-than-Nyquist signalling \cite{SE-MOCZ, ZF-MOCZ}. Sahin incorporated majority voting to 
enhance over-the-air computation \cite{OTA_MV}, and Perre et al. introduced smooshed binary MOCZ (SBMOCZ) to mitigate effects of timing offset \cite{SMOOSHED-MOCZ}.  

Index modulation (IM) is a well-established technique for improving bit error rate (BER) and SE by encoding additional information in the indicies of transmission resources, with IM having been applied to a number of waveforms, e.g., see \cite{AFFINE_IM} and the references therein. Despite its success in these waveforms, the integration of IM with MOCZ has not yet been explored. Motivated by the potential to further increase the SE of MOCZ, we propose IM-MOCZ, a scheme that encodes implicit bits through codebook selection while transmitting the remaining information using conventional binary-MOCZ (BMOCZ). At the receiver, Root Finding Minimum Distance (RFMD) or Direct Zero-Testing (DiZeT) detectors evaluate the received zeros across all candidate codebooks, and a majority vote process applied to the resulting penalty matrix is used to recover the transmitted message. Simulation results demonstrate that IM-MOCZ provides substantial SE improvements over conventional BMOCZ while also achieving improved BER and block error rate (BLER) performance for high values of $K$ compared to $N$.

This paper is organized as follows: Section \ref{sec:system_model} introduces the IM-MOCZ system model, while Section \ref{sec:proposed} outlines the design and integration of IM components. Simulation results are presented in Section \ref{sec:simulation} and Section \ref{sec:conclu} concludes the paper. 

\emph{Notations:} Throughout the paper, boldface lowercase and uppercase letters denote vectors and matrices, respectively. The notation $\mathbf{X}_{a,b}$ represents the elements (a,b) of matrix $\mathbf{X}$. We use $\mathbb{R}$ and $\mathbb{C}$ to represent the set of real and complex numbers, respectively. The symbol $*$ denotes the convolution operator. Conjugation of a complex value $c$ is denoted by $\bar{c}$. The transpose of a vector $\mathbf{a}$ is denoted by $\mathbf{a}^{\rm T}$. Binary numbers are represented using subscript $2$, e.g., $10110_2$.

\vspace{-0.5em}
\section{System Model}\label{sec:system_model}

Fig. \ref{fig:IM_MOCZ_BlockDiagram} presents the block diagram of the proposed IM-MOCZ communication system. The message to be transmitted, of length $N$, is divided into two bit streams. The first stream, consisting of the first $N-K$ bits ($b_1, \dots, b_{N-K}$), is used to select a unique Huffman BMOCZ codebook $\mathbf{\zeta}_i$, $i = 1, \dots, 2^{N-K}$, from a set of predefined codebooks $\mathcal{C}$ with radius $R$ using IM. 
The second bit stream consists of the remaining $K$ bits ($b_{N-K+1}, \dots, b_N$) transmitted using conventional BMOCZ according to the selected codebook $\mathbf{\zeta}_i$. In other words, each of the $K$ bits of the second stream is mapped to a unique complex zero $a_{i,k}\in \mathbb{C}$ or $1/\bar{a}_{i,k}\in \mathbb{C}$, $k = 1, \dots, K$, in the $z$-domain according to codebook $\mathbf{\zeta}_i$. A logical 0 is mapped to the inner complex zero $1/\bar{a}_{i,k}$, while a logical 1 is mapped to the outer complex zero $a_{i,k}$. Then, the $K+1$ polynomial coefficients $\mathbf{x}=[x_1, \dots, x_{k}, \dots, x_{K+1}]^{\rm T} \in \mathbb{C}^{K + 1}$, corresponding to the $K$ {transmitted}  complex zeros, are constructed according to the modified Toeplitz iterator method \cite{SE-MOCZ}. The proposed IM-MOCZ system transmits $N$ bits using $K+1$ ($K \leq N$) polynomial coefficients.

The received signal, $\mathbf{y} \in \mathbb{C}^{(K+L_{\rm{ch}})}$, is expressed as \cite{MOCZ}
\begin{equation}
    \mathbf{y} = \mathbf{x} * \mathbf{h} + \mathbf{w},
\end{equation}
where $\mathbf{h} \in \mathbb{C}^{L_{\text{ch}}}$ is the time-invariant multipath channel consisting of $L_{\rm{ch}}$ taps and $\mathbf{w} \in \mathbb{C}^{(K+L_{\rm{ch}})}$ is the additive white Gaussian noise (AWGN) with a one-sided power spectral density of $N_0$. The received sampled signal $\mathbf{y}$ is processed by a detection block employing RFMD or DiZeT detection schemes \cite{MOCZ} to estimate the complex zeros $\hat{a}_{i,k}$ or $1/\bar{\hat{a}}_{i,k}$ associated with each candidate codebook $\mathbf{\zeta}_i$. The resulting estimates are collected in the matrix $\hat{\mathbf{A}} \in \mathbb{C}^{2^{N-K} \times K}$, while the corresponding penalty metrics {(defined in Section \ref{sec:proposed})} for each estimated complex zero are stored in $\mathbf{P} \in \mathbb{R}^{2^{N-K} \times K}$. The transmitted message is recovered by identifying the most likely codebook $\mathbf{\hat{\zeta}}_{i}$ using a majority vote decision. The index of the recovered codebook determines the implicitly transmitted bits ($\hat{b}_1, \dots, \hat{b}_{N-K}$). The remaining explicitly transmitted bits are then recovered by selecting the detected complex zeros from $\hat{\mathbf{A}}$ corresponding to $\mathbf{\hat{\zeta}}_{i}$, yielding ($\hat{b}_{N-K+1}, \dots, \hat{b}_N$).

\section{Proposed IM-MOCZ}\label{sec:proposed}
As discussed in Section \ref{sec:system_model}, the transmission requires $K+1$ polynomial coefficients generated from the $K$ complex zeros $a_{i,k}$, $k = 1, ..., K$, corresponding to a specific codebook $\mathbf{\zeta}_i$. The key distinction between conventional BMOCZ and the proposed IM-MOCZ scheme lies in the handling of the first $N-K$ implicitly transmitted bits and the resulting codebook generation.

\subsection{IM-MOCZ Codebook Construction}

\begin{figure}[t]
    \centering
    {\includegraphics[width=0.3\textwidth]{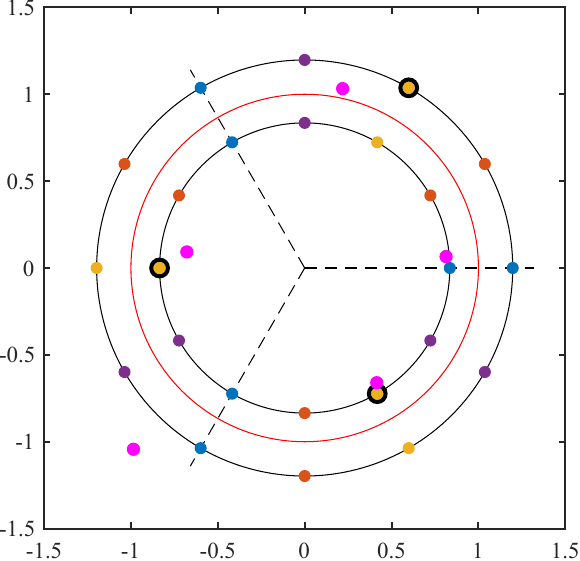}}
    \caption{{Complex zeros of the received signal for IM-MOCZ with $N=5$, $K=3$, $L_{\text{ch}}=3$, $E_b/N_0=10$~dB, $R=1.1974$, and message $10100_2$. Hollow black circles denote transmitted complex zeros of $\mathbf{x}$; magenta circles denote complex zeros of $\mathbf{y}$. Codebooks $\zeta_1$, $\zeta_2$, $\zeta_3$, $\zeta_4$ are shown in blue, orange, yellow, and purple, respectively. Received complex zeros are $\tilde{a}_1=0.8162\angle0.08059$, $\tilde{a}_2=1.0552\angle1.362$, $\tilde{a}_3=0.6846\angle3.006$, $\tilde{a}_4=1.435\angle-2.3277$, and $\tilde{a}_5=0.7796\angle-1.0098$.}}
    \vspace{-1em}
    \label{fig:IM-MOCZ_CodebookEx}
    
\end{figure}

In the proposed IM-MOCZ system, each candidate codebook $\mathbf{\zeta}_i$ is generated by applying a phase rotation to a common conventional Huffman BMOCZ codebook, denoted as $\mathbf{\zeta}_1$. Specifically, the $i$-th codebook is obtained by applying the phase offset
\begin{equation}
    \theta_i=\frac{2\pi(i-1)}{K2^{N-K}}, \quad i = 1,\dots, 2^{N-K},
\end{equation}
such that each complex zero location in the common conventional BMOCZ codebook is mapped according to
\begin{IEEEeqnarray}{RCL}
    \mathbf{\zeta}_i &{}={}&\mathbf{\zeta}_1e^{j\theta_i}, \quad i = 1,\dots, 2^{N-K}.
\end{IEEEeqnarray}
This produces $2^{N-K}$ phase-rotated codebooks that are uniformly spaced. Each codebook index $i$ is associated with a unique binary sequence. In this work, the sequence is defined by the $(N-K)$-bit binary representation of $i-1$, with the most significant bit corresponding to $b_1$. Selecting a codebook implicitly conveys the first $N-K$ information bits without increasing the number of polynomial coefficients transmitted. An example is illustrated in Fig. \ref{fig:IM-MOCZ_CodebookEx}. In addition to the angular placement of the zeros, the radial placement also plays an important role in IM-MOCZ performance. Following \cite{MOCZ}, the optimal radius is chosen as $R = \sqrt{1+2\lambda\sin(\pi/K)}$, where $\lambda$ balances the radial and phase separation of the codebook zeros. This dependence on $K$ is particularly important in IM-MOCZ, since varying $K$ directly affects the complex zero distribution across sectors.

\subsection{Per-Sector Detection and Penalty Metrics}

For each candidate codebook $\mathbf{\zeta}_i$ and sector $k \in \{1,\dots,K\}$, the detector produces an estimated complex zero $\hat{a}_{i,k}\in \mathbb{C}$ or $1/\bar{\hat{a}}_{i,k}\in \mathbb{C}$ together with an associated penalty metric $p_{i,k} \in \mathbb{R}$. The penalty metric quantifies how well the received polynomial zeros match the expected codebook zero in a given sector. In this work, two detection schemes are considered: RFMD and DiZeT detection \cite{MOCZ}. A discussion on RFMD is included to show that IM-MOCZ can be applied with multiple detection strategies. The received polynomial $\mathbf{y}$ produces $M = K + L_{\text{ch}}-1$ complex zeros, denoted $\tilde{a}_m$, $m=1, \dots, M$.

\subsubsection{RFMD Penalty Metric}
For RFMD detection, a detector block is applied to each candidate codebook $\mathbf{\zeta}_i$ across all $K$ sectors. Given the received polynomial coefficients $\mathbf{y}$ and codebook $\mathbf{\zeta}_i$, the RFMD block returns, for each sector $k$, a detected complex zero $\hat{\mathbf{A}}_{i,k}$ together with an associated penalty metric $p_{i,k}^{\text{(RFMD)}} = \mathrm{d}(\tilde{a}_m, \hat{\mathbf{A}}_{i,k})$, where $\mathrm{d}(\tilde{a}_m, \hat{\mathbf{A}}_{i,k})$ corresponds to the smallest Euclidean distance between the received complex zero $\tilde{a}_m$ in sector $k$ of $\mathbf{\zeta}_i$ and $\hat{\mathbf{A}}_{i,k} \in \{a_{i,k}, 1/\bar{a}_{i,k}\}$ which is the codebook complex zero that achieves this minimum. This operation is expressed as
\begin{equation}
    \big(\hat{\mathbf{A}}_{i,k},\, p_{i,k}^{\text{(RFMD)}} \big)
    = \text{RFMD}(\mathbf{y}, \mathbf{\zeta}_i).
\end{equation}
Smaller values of $p_{i,k}^{\text{(RFMD)}}$ indicate a closer match between the received and detected codebook complex zeros, and therefore a higher confidence in detection. These penalty metrics are used directly in the majority vote codebook detection procedure.

\subsubsection{DiZeT Penalty Metric}
For DiZeT detection, a detector block is applied to each candidate codebook $\mathbf{\zeta}_i$ across all $K$ sectors. Given the received polynomial coefficients $\mathbf{y}$ and a candidate codebook $\mathbf{\zeta}_i$, the DiZeT detector evaluates, for each sector $k$, a detected complex zero $\hat{\mathbf{A}}_{i,k}$ and corresponding penalty metric $p^{\text{(DiZeT)}}_{i,k}$\footnote{For a more efficient implementation of $p^{\text{(DiZeT)}}_{i,k}$ for rotated versions of the standard MOCZ codebook $\zeta_i$, the oversampled IDFT-based DiZeT decoder in \cite{MOCZ-Prac} can be used. Moreover, $p^{\text{(DiZeT)}}_{i,k}$ is equivalent to the $\min \{\cdot\}$ term of \cite[(34)]{MOCZ-Prac}.}. This is achieved by substituing the codebook zero locations associated with $\zeta_{i}$ into the received polynomial formed from the received complex zeros $\tilde{a}_m$ \cite{MOCZ}. This is expressed as

\begin{equation}
    \big(\hat{\mathbf{A}}_{i,k},\, p_{i,k}^{\text{(DiZeT)}} \big)
    = \text{DiZeT}(\mathbf{y}, \mathbf{\zeta}_i).
\end{equation}
During detection, the DiZeT detector must determine whether the received zero corresponds to the inner or outer zero of the codebook pair $\{a_{i,k}, 1/\bar{a}_{i,k}\}$. Accordingly, two candidate penalty metrics are computed: an outer penalty $p^{\text{(out)}}_{i,k}$ and an inner penalty $p^{\text{(in)}}_{i,k}$ \cite{MOCZ}. These are defined as

\begin{IEEEeqnarray}{RCL}
    \label{eqn:DiZeT_products_IM}
    p_{i,k}^{\text{(out)}} &=& 
    \Bigg|\prod_{m=1}^{M} (a_{i,k} - \tilde{a}_m) \Bigg|, \\
    p_{i,k}^{\text{(in)}} &=& 
    R^{M} \Bigg|\prod_{m=1}^{M} (a_{i,k}^{-1} - \tilde{a}_m) \Bigg|,
\end{IEEEeqnarray}

\noindent where $R$ denotes the radius of the outer codebook zeros. Smaller penalty values indicate a closer agreement between the received complex zeros and the associated codebook complex zero locations, and therefore higher confidence in detection.

The DiZeT penalty metric for codebook $\mathbf{\zeta}_i$ in sector $k$ is then defined as the minimum of the inner and outer penalties,
\begin{equation}
    \label{eqn:DiZeT_penalty}
    p_{i,k}^{\text{(DiZeT)}} = \min\big(p^{\text{(in)}}_{i,k}, p_{i,k}^{\text{(out)}}\big).
\end{equation}
In the event that $p^{\text{(in)}}_{i,k}=p^{\text{(out)}}_{i,k}$, the DiZeT penalty metric is selected as either value with equal probability, and the detected complex zero $\hat{\mathbf{A}}_{i,k}$ is selected as $1/\bar{a}_{i,k}$ or $a_{i,k}$, respectively.

\textbf{Example 1}: 
In Fig. \ref{fig:IM-MOCZ_CodebookEx}, consider sector $k = 1$ and codebook $i = 1$ (with associated complex zero pair $\{a_{1,1}, 1/\bar{a}_{1,1}\} = \{1.1974,0.8351\})$.
The outer and inner DiZeT penalties are calculated as
\vspace{-0.1em}
\begin{align}
    p_{1,1}^{\text{(out)}} &= 
    \Big|(1.1974 - \tilde{a}_1)(1.1974 - \tilde{a}_2)(1.1974 - \tilde{a}_3) \\
    &\quad \cdot (1.1974 - \tilde{a}_4)(1.1974 - \tilde{a}_5)\Big| 
    = 2.5766, \nonumber\\
    p_{1,1}^{\text{(in)}} &= 
    (1.1974)^5 \cdot \Big| (1/\bar{a}_{1,1} - \tilde{a}_1) \cdots (1/\bar{a}_{1,1} - \tilde{a}_5) \Big| 
    = 0.5095.
\end{align}
Comparing the two penalties, the DiZeT detector selects the smaller magnitude. In this case, $p_{1,1}^{\text{(in)}} = 0.5095$ is smaller than $p_{1,1}^{\text{(out)}} = 2.5766$. Therefore, the detector chooses the corresponding inner zero of that sector and outputs
\[
    \hat{\mathbf{A}}_{1,1} = 1/\bar{a}_{1,1}, \quad
    p_{1,1}^{\text{(DiZeT)}} = \min(p_{1,1}^{\text{(in)}}, p_{1,1}^{\text{(out)}}) = 0.5095.
\]
That is, the value of $\hat{\mathbf{A}}_{1,1}$ is determined by which penalty metric is smaller, selecting either the inner or outer zero of the corresponding sector.

Repeating this procedure for all sectors and codebooks, the resulting penalty matrix $\mathbf{P}^{\text{(DiZeT)}}$ and the detected zeros matrix $\hat{\mathbf{A}}^{\text{(DiZeT)}}$ are, respectively, calculated as:

\begin{IEEEeqnarray}{RCLr}
    \mathbf{P}^{\text{(DiZeT)}} &=&
    \begin{bmatrix} \label{eqn:DiZeT_Penalty_Matrix}
        {0.5095} & {4.9687} & {1.8879}\\
        {2.5785} & {3.3317} & {2.9278}\\
        {2.6803} & {1.6255} & {0.4769}\\
        {2.3067} & {2.0498} & {1.9786}
    \end{bmatrix}, \\
    \hat{\mathbf{A}}^{\text{(DiZeT)}} &=&
    \begin{bmatrix}
        {1/\bar{a}_{1,1}} & {1/\bar{a}_{1,2}} & {a_{1,3}}\\
        {1/\bar{a}_{2,1}} & {1/\bar{a}_{2,2}} & {1/\bar{a}_{2,3}}\\
        a_{3,1} & {1/\bar{a}_{3,2}} & {1/\bar{a}_{3,3}}\\
        a_{4,1} & {a_{4,2}} & {1/\bar{a}_{4,3}}
    \end{bmatrix}. &  \hspace{1cm}\blacksquare
\end{IEEEeqnarray}

\subsection{Majority Vote Codebook Detection}

Once all penalty values 
\(p_{i,k} \in \{p_{i,k}^{\text{(RFMD)}}, p_{i,k}^{\text{(DiZeT)}}\}\) 
have been computed for every candidate codebook \(\mathbf{\zeta}_i\) and collected in the penalty matrix \(\mathbf{P} \in \{\mathbf{P}^{\text{(RFMD)}}, \mathbf{P}^{\text{(DiZeT)}}\}\), the final codebook $\hat{\zeta}_i$ is detected using a majority vote across sectors. 


For each sector \(k\), the codebook index that minimizes the penalty metric is determined as
\begin{equation}
    \hat{i}_k = \arg\min_i p_{i,k}.
\end{equation}
Each sector casts one vote for the codebook that minimizes the penalty in that sector.  
The total number of votes for each candidate codebook is given by the number of sectors that select it as the minimum-penalty codebook:
\begin{equation}
    w_i = \left| \{ k \in \{1,\dots,K\} \mid \hat{i}_k = i \} \right|.
\end{equation}
The final codebook decision is made by selecting the codebook with the majority of votes:
\begin{equation}
    \hat{i} = \arg\max_i w_i.
\end{equation}

In the event of a tie, the detected codebook is chosen uniformly at random from the set of tied candidates.  
Once \(\mathbf{\hat{\zeta}}_{i}\) is selected, the implicitly transmitted bits 
\((\hat{b}_1, \dots, \hat{b}_{N-K})\) are recovered directly from \(\hat{i}\). 
The explicitly transmitted bits 
\((\hat{b}_{N-K+1}, \dots, \hat{b}_N)\) are then recovered by selecting the detected zeros \(\hat{a}_{i,k}\) associated with \(\mathbf{\hat{\zeta}}_{i}\) from the matrix \(\hat{\mathbf{A}}\) and applying the standard MOCZ demapping procedure.

\textbf{Example 2}: 
Consider the DiZeT penalty matrix $\mathbf{P}^{\text{(DiZeT)}}$ in \eqref{eqn:DiZeT_Penalty_Matrix}, where
each row corresponds to a candidate codebook and each column corresponds to a sector.
For each sector \(k\), we find the codebook index \(\hat{i}_k\) that yields the minimum penalty:

\begin{equation}
    \begin{aligned}
    \min\{0.5095, 2.5785, 2.6803, 2.3067\} = 0.5095 \Rightarrow \hat{i}_1 = 1,\\
    \min\{4.9687, 3.3317, 1.6255, 2.0498\} = 1.6255 \Rightarrow \hat{i}_2 = 3,\\
    \min\{1.8879, 2.9278, 0.4769, 1.9786\} = 0.4769 \Rightarrow \hat{i}_3 = 3.\\
    \end{aligned}
\end{equation}
For each candidate codebook $\mathbf{\zeta}_i$, collect the sectors that voted for it (i.e., sectors in which codebook $\mathbf{\zeta}_i$ achieves the minimum penalty).  
The number of votes \(w_i\) is then taken as the cardinality of the set:

\begin{equation}
    \begin{alignedat}{3}
        w_1 & = |\{1\}|       & = 1, \quad & 
        w_2 & = |\{\}|        & = 0, \\
        w_3 & = |\{2,3\}|     & = 2, \quad & 
        w_4 & = |\{\}|        & = 0,
    \end{alignedat}
\end{equation}
and the codebook with the most votes is then selected:
\[
    \hat{i} = \arg\max_i w_i = 3.
\]
The detected zeros for the selected codebook are extracted from the RFMD detected zero matrix as $\hat{\mathbf{A}}_{3} = 
\begin{bmatrix}
\hat{a}_{3,1} &1/\bar{\hat{a}}_{3,2} & 1/\bar{\hat{a}}_{3,3}
\end{bmatrix}.$

The implicitly transmitted bits \((\hat{b}_1, \hat{b}_{2})\) are recovered directly from the codebook index \(\hat{i}-1 = 2\) as $10_2$. The explicitly transmitted bits \((\hat{b}_{3}, \hat{b}_{4}, \hat{b}_{5})\) are recovered as $100_2$ from the detected zeros \(\hat{\mathbf{A}}_{{3}}\) using the standard BMOCZ demapping procedure. The final result is concatenated as $10100_2$. $\hfill\blacksquare$

\subsection{Complexity Analysis of IM-MOCZ}

The IM-MOCZ receiver operates as an additional processing layer on top of a 
conventional BMOCZ detector, so its complexity is analyzed in terms of the overhead 
introduced relative to a single detection. Let $V \in \{V^{\text{(RFMD)}}, 
V^{\text{(DiZeT)}}\}$ denote the complexity of one such detection, which depends on 
the chosen detector implementation.

Since IM-MOCZ evaluates all $2^{N-K}$ candidate codebooks ${\zeta}_i$, 
the detection stage requires $2^{N-K}$ independent detections, contributing 
$O(2^{N-K} V)$. The majority voting stage then determines, for each of the $K$ physically transmitted bits, the most frequent decision across all $2^{N-K}$ outputs, requiring $O(2^{N-K} K)$ comparisons. The final indexing and demapping steps are $O(K)$ and are dominated by the preceding terms.

The overall complexity is therefore $O\left(2^{N-K}(V + K)\right)$,
reflecting an exponential overhead of $2^{N-K}$ relative to conventional BMOCZ 
detection, which requires only $O(V)$. For the SPC setting considered in this work, 
$N - K$ can be kept small, ensuring that the added complexity 
remains manageable regardless of the underlying detector used.

\vspace{-0.2em}
\section{Simulation Results}\label{sec:simulation}
This section evaluates the BER performance of the proposed IM-MOCZ scheme and compares it with conventional BMOCZ employing RFMD and DiZeT detection. Simulations use the equivalent baseband model shown in Fig.~\ref{fig:IM_MOCZ_BlockDiagram}, with the channel normalized to have an average energy of one with a flat power delay profile. The number of Monte Carlo trials used in the simulation depend on the value of $E_b/N_0$. Message length for conventional BMOCZ is denoted as $N_{\mathrm{M}}$. To have a fair comparison between IM-MOCZ and MOCZ, the total transmit energy of both schemes are set to be equal to $N + L_{\text{ch}}$ joules. Accordingly, $E_b/N_0 = (N_{\mathrm{M}} + L_{\text{ch}})/(N_0 N_{\mathrm{M}})$ for MOCZ and $E_b/N_0 = (N + L_{\text{ch}})/(N_0 K)$ for IM-MOCZ.

Two simulation scenarios are considered. In the first, message lengths are fixed to be equal ($N_{\mathrm{M}}=N$). This scenario reflects practical SPC scenarios where transmissions are infrequent and payloads are small. In the second scenario, a fair comparison is made under equal SE, measured in bits/s/Hz. The spectral efficiencies are given by $\mathrm{SE}_{\mathrm{M}} = N_{\mathrm{M}}/(N_{\mathrm{M}} + L_{\text{ch}})$ for conventional BMOCZ and $\mathrm{SE} = N/(K + L_{\text{ch}})$ for IM-MOCZ.

\subsection{Equal Transmit Bits}\label{ssec:Equal_Transmit}
Figure~\cref{fig:Equal_Tx} presents the BER and BLER performance of IM-MOCZ using RFMD and DiZeT detection for different values of $K$, where $N = N_{\mathrm{M}} = 10$ and $L_{\text{ch}} = 3$, compared to conventional BMOCZ ($N_{\mathrm{M}}=K=10$). For $K = 8$, IM-MOCZ achieves a BER gain of approximately 0.2~dB and 0.9~dB at a BER of $10^{-4}$ using RFMD and DiZeT detectors respectively. For $K = 6$, the gain increases to approximately 2.1~dB using RFMD and 2.3~dB using DiZeT detectors. Reducing $K$ increases the energy per transmitted bit, generally improving BER performance. However, smaller values of $K$ also increase the number of implicit bits ($N-K$), resulting in a larger set of candidate codebooks (e.g., $2^{10-4} = 64$). Consequently, a single error in detecting transmitted bits can propagate to all $N-K$ implicit bits, which significantly degrades BER and BLER performance, as observed when $K = 4$. Under equal transmit energy, conventional BMOCZ achieves $\mathrm{SE}_{\mathrm{M}} = 10/13$ bits/s/Hz, while IM-MOCZ achieves $\mathrm{SE} = 10/11$, $10/9$, and $10/7$ bits/s/Hz for $K = 8, 6, 4$, respectively, demonstrating that IM-MOCZ can attain SE higher than $1$ bits/s/Hz, unlike conventional BMOCZ.

\subsection{Equal SE}

A fair comparison under equal SE is meaningful only for $\mathrm{SE} \le 1$, since for conventional BMOCZ $N_{\mathrm{M}} < N_{\mathrm{M}}+L_{\mathrm{ch}}$. In contrast, IM-MOCZ enables transmission beyond this limit by encoding additional information through index modulation, effectively enabling $N > K+L_{\mathrm{ch}}$. Therefore, comparisons for $\mathrm{SE} > 1$ can not be applied to conventional BMOCZ and are instead used to highlight the extended operating range of IM-MOCZ.

\begin{figure*}[ht]
\centering
\newcommand{\figw}{0.304\textwidth}
\begin{minipage}[t]{0.627\textwidth}
    \centering
    \includegraphics[width=0.485\linewidth]{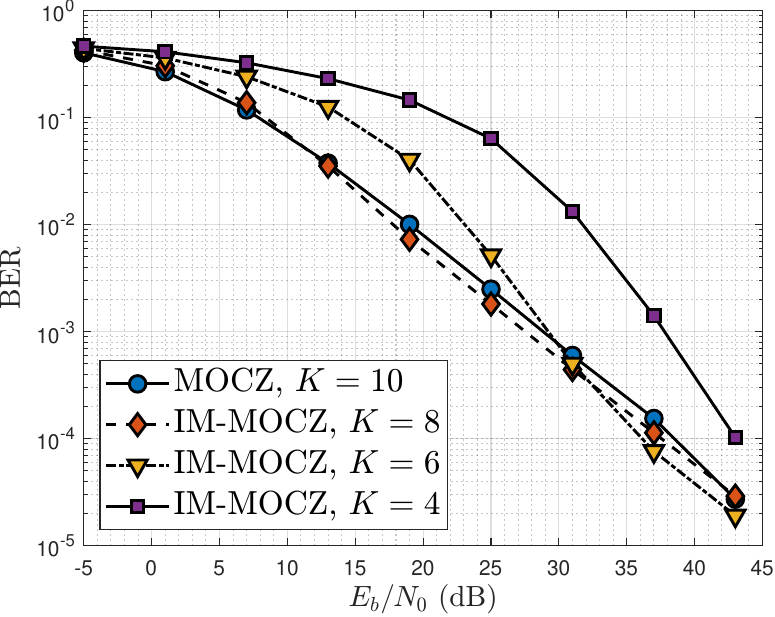}
    \hfill
    \includegraphics[width=0.485\linewidth]{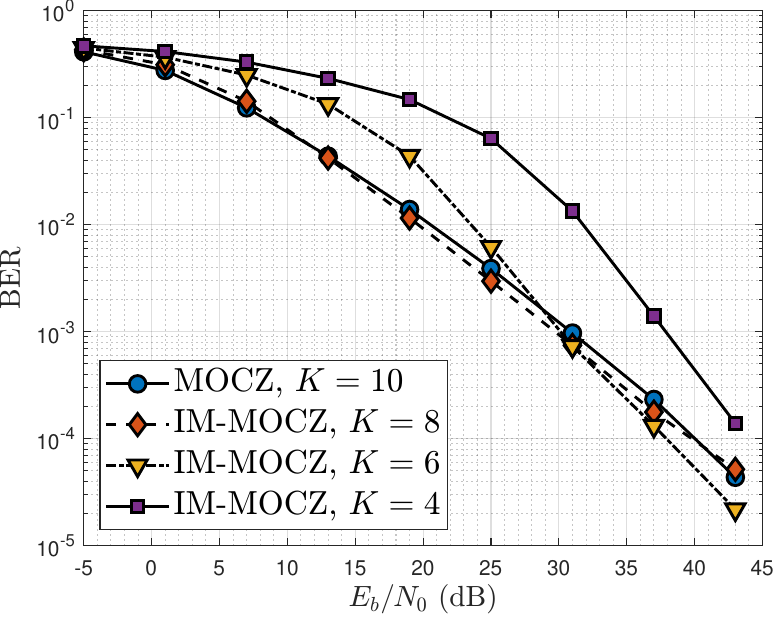}
    \vspace{0.3em}
    \includegraphics[width=0.485\linewidth]{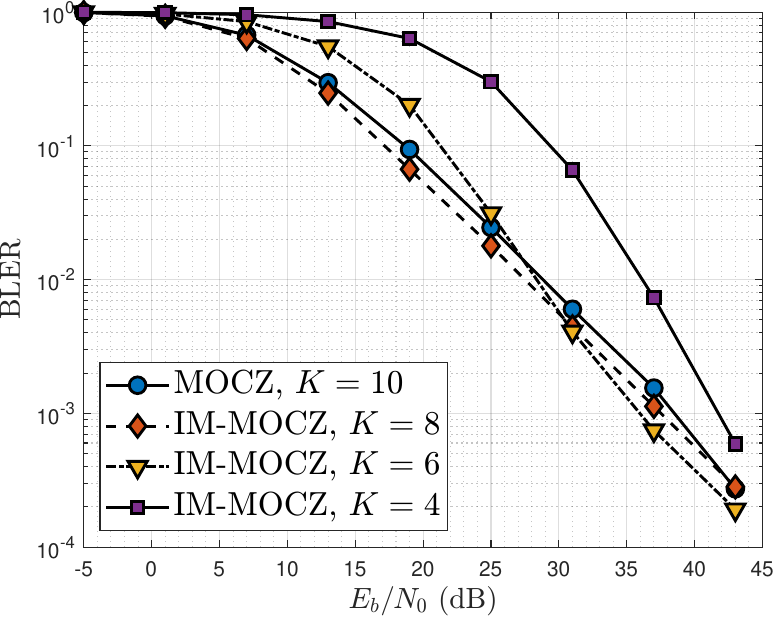}
    \hfill
    \includegraphics[width=0.485\linewidth]{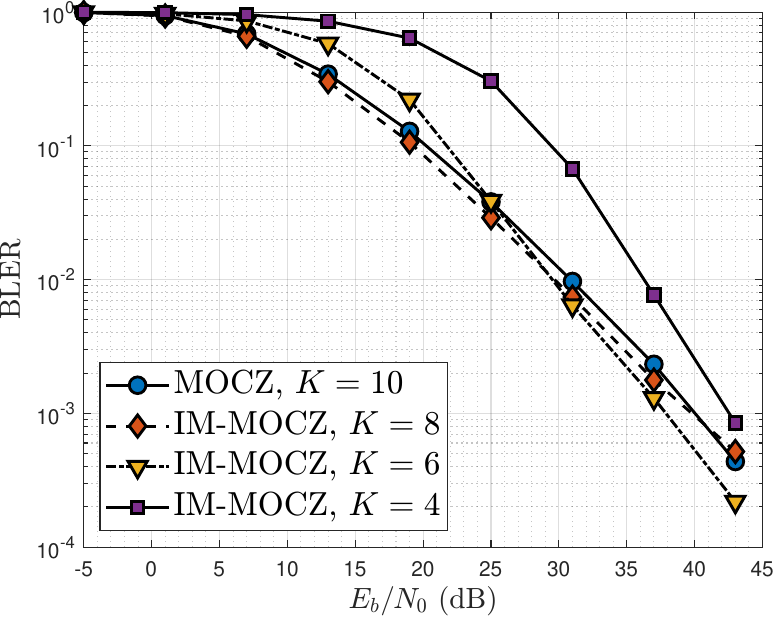}
    \captionof{figure}{Equal Tx bits BER (top) and BLER (bottom) using DiZeT (left) and RFMD (right) detectors with $N=N_{\mathrm{M}}=10$, and $L_{\text{ch}}=3$.}
    \vspace{-1em}
    \label{fig:Equal_Tx}
\end{minipage}
\hspace{0.012\textwidth}
\begin{minipage}[t]{0.304\textwidth}
    \centering
    \includegraphics[width=\linewidth]{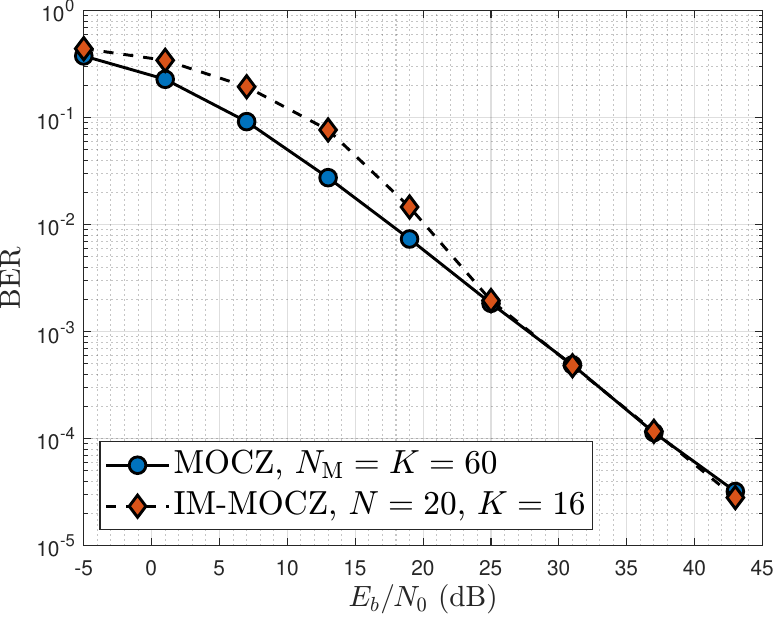}
    \vspace{0.3em}
    \includegraphics[width=\linewidth]{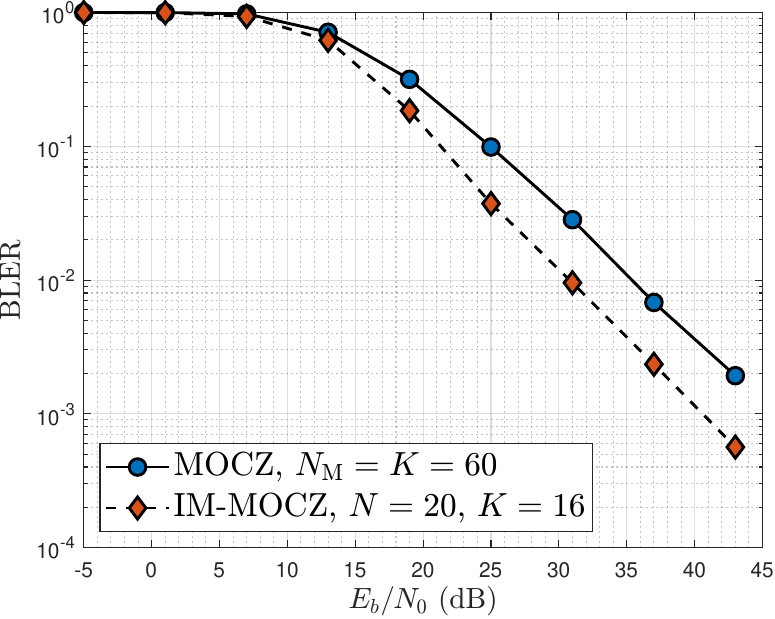}
    \captionof{figure}{Equal SE BER (top) and BLER (bottom) using DiZeT detector with $L_{\mathrm{ch}}=6$.}
    \label{fig:equal_SE}
    \vspace{-1em}
\end{minipage}
\end{figure*}

Figure~\cref{fig:equal_SE} presents BER and BLER results under equal SE using the DiZeT detector. For IM-MOCZ, parameters $N = 20$, $L_{\mathrm{ch}} = 6$, and $K = 16$ are used, yielding $\mathrm{SE} = 10/11$ bits/s/Hz. To match this SE, conventional BMOCZ requires a significantly longer message length of $N_{\mathrm{M}} = 60$ with $L_{\mathrm{ch}} = 6$.

Under equal SE, both schemes achieve comparable BER performance over most $E_b/N_0$ values. IM-MOCZ shows slightly higher BER at low $E_b/N_0$, due to increased sensitivity in detecting implicit bits. At high $E_b/N_0$, however, IM-MOCZ is disadvantaged when comparing BER. This is because 
matching SE forces conventional BMOCZ to use a significantly longer block ($N_{\mathrm{M}} = 60$ vs. $N = 20$), which dilutes the relative impact of mis-detections due to channel zeros across more bits. At high $E_b/N_0$, noise is negligible and mis-detection due to channel zeros dominates; conventional BMOCZ benefits from its longer block length, as the fixed number of channel-induced errors represents a smaller fraction of the total transmitted bits compared to IM-MOCZ. In contrast, BLER performance consistently favors IM-MOCZ, as conventional BMOCZ requires longer blocks to satisfy the same SE, increasing the probability that at least one bit error occurs within a block.

\vspace{-0.5em}
\section{Conclusion} \label{sec:conclu}
In this paper, we incorporate IM into the non-coherent MOCZ scheme, conveying additional information via a Huffman BMOCZ codebook index and improving achievable SE. Under equal transmitted bits, IM-MOCZ with $N=10$, $L_{\mathrm{ch}}=3$, and $K=6$ achieves a 44\% SE gain and up to 2.3~dB BER improvement over conventional BMOCZ using the DiZeT detector, with gains most pronounced at high $E_b/N_0$, where reduced noise enables more reliable codebook detection. Smaller values of $K$ enable $\mathrm{SE}\ge 1$ unattainable by conventional BMOCZ, but increase detection complexity exponentially, while beyond a certain value of $K$, BER performance degrades. Under equal SE, IM-MOCZ achieves comparable BER across most $E_b/N_0$ values, though conventional BMOCZ gains a BER advantage at high $E_b/N_0$ due to longer blocks diluting channel-zero-induced errors, while BLER consistently favors IM-MOCZ. Overall, IM-MOCZ offers a trade-off between SE, and detection complexity, making it a promising candidate for SPC scenarios requiring non-coherent operation and improved spectral efficiency.

\vspace{-0.1em}
\bibliographystyle{IEEEtran}
\bibliography{IEEEabrv,References}

\end{document}